\documentclass[sigconf, authorversion]{acmart}

\usepackage{tikz}
\usepackage{colortbl}
\usepackage{enumitem}
\usepackage{multicol}
\usepackage{multirow}
\usepackage{pgfplots}
\usepackage{pgfplotstable}
\usepackage{quoting}
\usepackage{subcaption}
\usepackage{tcolorbox}
\usepackage{verbatim}
\usepackage{balance}

\pgfplotsset{compat=1.18}
\usetikzlibrary{patterns}

\definecolor{good}{rgb}{0.267004, 0.004874, 0.329415}   
\definecolor{okay}{rgb}{0.190631, 0.407061, 0.556089}   
\definecolor{notgood}{rgb}{0.368627, 0.788888, 0.482758} 
\definecolor{bad}{rgb}{0.993248, 0.906157, 0.143936}    
\definecolor{none}{rgb}{1, 1, 1}
\AtBeginDocument{%
  }

\setcopyright{acmlicensed}
\copyrightyear{2018}
\acmYear{2018}
\acmDOI{XXXXXXX.XXXXXXX}

\acmConference[Conference acronym 'XX]{Make sure to enter the correct
  conference title from your rights confirmation emai}{June 03--05,
  2018}{Woodstock, NY}
\acmISBN{978-1-4503-XXXX-X/18/06}

\begin{document}

\title[EiPL Questions with Indic Languages]{Explain in Plain Language Questions with Indic Languages: Drawbacks, Affordances, and Opportunities}

\author{David H Smith IV}
\orcid{0000-0002-6572-4347}
\affiliation{
  \institution{University of Illinois}
  \city{Urbana}
  \state{IL}
  \country{USA}
}
\email{dhsmith2@illinois.edu}

\author{Viraj Kumar}
\orcid{0000-0002-2252-0141}
\affiliation{
  \institution{Indian Institute of Science}
  \city{Bengaluru}
  \state{Karnataka}
  \country{India}
}
\email{viraj@iisc.ac.in}

\author{Paul Denny}
\orcid{0000-0002-5150-9806}
\affiliation{
  \institution{University of Auckland}
  \city{Auckland}
  \country{New Zealand}
}
\email{p.denny@auckland.ac.nz}

\renewcommand{\shortauthors}{David H. Smith IV, Viraj Kumar, and Paul Denny}

\begin{abstract}

    \textbf{Background:} Introductory computer science courses use
    ``Explain in Plain English'' (EiPE) activities to develop and assess
    students' code comprehension skills, but creating effective autograders for
    these questions is challenging and limited to English. This is a particular
    challenge in linguistically diverse countries like India where students may
    have limited proficiency in English.

    \noindent\textbf{Objective:} We evaluate the efficacy of a recently
    introduced approach called Code Generation Based Grading (CGBG) in enabling
    language agnostic ``Explain in Plain Language'' (EiPL) activities. Here
    students' EiPL responses generate code that is tested for functional
    equivalence to the original which was being described.

    \noindent\textbf{Method:} We initially evaluate the correctness of code generated
    from correct EiPL responses provided in 10 of India's
    most commonly spoken languages. To evaluate the effectiveness of the
    approach in practice, we assess student success and perceptions of EiPL
    questions in a NPTEL (National Programme on Technology Enhanced Learning)
    course.

    \noindent\textbf{Results:} We find promising results for the correctness of
    code generated from translations of correct EiPL responses, with most
    languages achieving a correctness rate of 75\% or higher. However, in
    practice, many students preferred to respond in English due to greater
    familiarity with English as a technical language, difficulties writing in
    their native language, and perceptions of the grader being less capable of
    generating code from prompts in their mother tongue.
    
\end{abstract}



\maketitle              
\section{Introduction}

Among the many skills introductory computer science students are expected to
acquire are the abilities to read code and comprehend its
purpose~\cite{izu2019fostering,xie2019theory,fowler2021autograding}. To develop
and evaluate these skills, ``Explain in Plain English'' questions, where students
are shown a segment of code and asked to explain it, are often
used~\cite{murphy2012ability,murphy2012explain}. Informed by the Structure of
the Observed Learning Outcome (SOLO) taxonomy~\cite{biggs2014evaluating},
responses are not only graded on their correctness but also whether they
describe the code's purpose (a \textit{relational} response) or simply it's
implementation (a \textit{multistructural} response)~\cite{lister2006not}.

To enable the use of these questions at scale, autograders have been developed
and achieved remarkable success~\cite{fowler2021autograding,azad2020strategies}.
However, one shortcoming of the existing approaches is that they rely on a large
quantity of human labeled data. Extending these approaches to a variety of
languages would require a corpus of labeled data to be created for each
language. A more recent autograding approach referred to as  ``Code Generation
Based Grading'' (CGBG) uses a Large Language Model (LLM) to generate code from a
student's EiPE response and grades the generated code to check if it matches the
functionality of the code the student was attempting to
describe~\cite{smith2023code,denny2024explaining,smith2024prompting}. Though
this approach currently has the shortcoming of not being able to distinguish
between \textit{multistructural} and \textit{relational} responses, it comes with
the benefits of being able to provide students more objective and specific feedback (e.g.,
generated code, test case results). In addition, it enables students to respond
to questions in languages other than English, transforming what was formerly an
``Explain in Plain English'' question into an ``Explain in Plain Language''
(EiPL) question~\cite{denny2024explaining,smith2024prompting}.

To explore the degree to which a state-of-the-art LLM (OpenAI's GPT-4o) is
language agnostic, and by extension, the degree to which CGBG enables EiPL
questions, in this paper we address the following research questions.\\
\begin{enumerate}[leftmargin=*, label=\textbf{RQ\arabic*:}]
    \item How effective is the CGBG process for EiPL questions in the context of India's linguistically diverse landscape?
    \item How do students interact with EiPL questions and what are their perceptions of responding in their mother tongue?
\end{enumerate}
As noted by other authors~\cite{kumar2021refute}, limited English language
proficiency among Indian students presents a notable barrier to the adoption of
code comprehension activities such as EiPE questions. CGBG presents a unique
possibility to address existing inequities in the assessment of code
comprehension by potentially allowing students to respond in their mother tongue
as well as other languages in which they have fluency.

\section{Background}

In this section, we present literature on code comprehension activities,
(Section~\ref{sec:code-comprehension}), the challenges of developing
autograders for these activities, (Section~\ref{sec:distractor-blocks}), and
how new autograding approaches and their affordances may be particularly
beneficial in the context of India's linguistic diversity
(Section~\ref{sec:irt}; Figure~\ref{fig:languages-by-state}).

\subsection{Code Reading and Comprehension Activities}\label{sec:code-comprehension}

Learning to program is often discussed as mastering a variety of skills, namely,
learning to read, write, and trace the execution of
code~\cite{xie2019theory,fowler2022reevaluating}. As is the case with mastering
any skill, frequent practice along with timely and actionable feedback are
essential~\cite{bangert1991instructional,hao2022towards}. Questions which
address code writing and tracing are relatively easy to develop and deploy, even
at scale~\cite{west2021integrating}. However, code reading questions, which
often involve students providing a natural language description of code, are
more difficult to grade automatically~\cite{azad2020strategies}. This in turn
limits their scalability, the variety of questions students have the opportunity
to practice with, and the timeliness of the feedback students receive. This is
particularly concerning in the age of AI code generation, where courses using
generative AI tools must emphasize these skills to ensure students can validate
the code they generate
~\cite{prather2023robots,vadaparty2024cs1,denny2024prompt}.

Perhaps the most common code reading activity is the ``Explain in Plain
English'' (EiPE) question~\cite{murphy2012explain,murphy2012ability}. In these
questions students are shown a segment of code and asked to provide an English
description of the code's purpose rather than it's implementation. Fowler et al.
found that instructors who include such questions in their courses particularly
value them for developing and evaluating their students' ability to reason about
the purpose of code \cite{fowler2021how}. Autograders, which rely on a large
corpus of human labeled training data, have been developed and perform similarly
to that of a trained teaching
assistant~\cite{fowler2021autograding,azad2020strategies}. Though these results
are promising, they are limited to dichotomous feedback which can limit the
ability of students to learn from that feedback~\cite{bangert1991instructional}
and can lead to distrust of the autograder~\cite{li2023wrong}.

Beyond the limitations relating to scale and the difficulties relating to
autograders, EiPE questions suffer from a singular limitation: only accepting
responses in English. All prior work, both in developing rubrics and creating
autograders, has been done exclusively in the context of English language
responses. To address this issue, Kumar \cite{kumar2021refute} introduced
\textit{Refute} questions, where students are shown a segment of code and a
stated task that it fails to achieve. To answer the question correctly the
student must select an input which demonstrates that failure. The broader goal
of these questions is to address the skills of task comprehension and code
comprehension in a single activity that is not dependent on a student's ability
to articulate the purpose of a given segment of code or the nature of an error
in natural language~\cite{agarwal2023bug}.

\subsection{Explain in Plain Language Questions}\label{sec:distractor-blocks}

As previously noted, the scalability of EiPE questions is limited by the
complexity of designing an autograder and the fact that autograders must be
created on a per question basis. In theory, if multiple languages were to be
supported, that would require the training of additional models for each
language making the method largely intractable using previous approaches. A more
recent approach to grading EiPE questions was developed by Smith IV et al.
\cite{smith2023code}, referred to as ``Code Generation Based Grading'' (CGBG).
This approach involves the use of LLMs to generate code from a student's EiPE
response and then tests that code's functionality using unit tests as a proxy
for grading the correctness of a student's response. When Denny et al.
\cite{denny2024explaining} and Smith IV et al. \cite{smith2024prompting}
incorporated the grading approach into a variety of CS1 lab activities they
found that several students successfully responded to these questions in
Chinese. This suggests the CGBG approach can transform the ``Explain in Plain
English'' activity into an ``Explain in Plain Language'' (EiPL) activity. It
must be noted that the approach is only as language agnostic at the LLM that
supports it and model performance will inevitably differ amongst
languages~\cite{zhang2023don,zhang2024m3exam,zhao2024large}. Nonetheless, this
finding represents both a promising step forward and an area ripe for
investigation, especially as the computing education community grapples for
productive ways to incorporate LLMs into the curriculum and classroom
practice~\cite{denny2024prompt,denny2024computing}.

\subsection{Language Diversity in India and Computer Science Education}\label{sec:irt}

\begin{figure*}
    \centering
    \includegraphics[width=\textwidth]{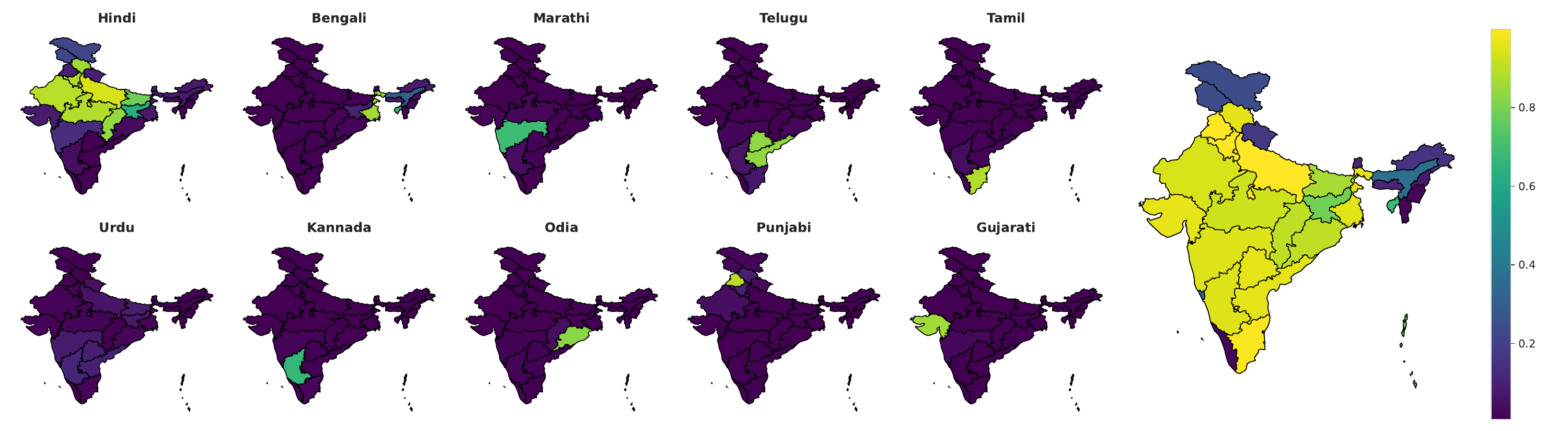}
    \caption{The percentage of each state's population which speaks the given language as their mother tongue as reported by the 2011 census for the top 10 most spoken languages of India's 22 scheduled languages. It is important to note that this census predates the separation of the states of Ladakh and Telangana from the states of Jammu-Kashmir and Andhra Pradesh, respectively. As such, it may not be fully representative of the languages spoken in those states today.}
    \label{fig:languages-by-state}
\end{figure*}

As India's National Education Policy (NEP 2020)~\cite{policy2020ministry} notes,
\begin{quote}
    ``\textit{The aim must be for India to have an education system by 2040 that is second to none, with equitable access to the highest-quality education for all learners regardless of social or economic background.}'' (p. 3)
\end{quote}
Underlying this goal, the policy states \textit{promoting multilingualism and
the power of language}, \textit{respect for local context}, and \textit{respect
for diversity} as being among its key principles. Though each of these are
certainly laudable goals, multilingualism in particular serves as a barrier to
the creation, adoption, and evaluation of curriculum given India's linguistic
diversity (\autoref{fig:languages-by-state}). 

In the context of computer science, and perhaps Science, Technology, Engineering
and Mathematics (STEM) courses more generally, we are fortunate than many of our
core instructional materials are language agnostic. Though instructional
materials, documentation, and question prompts must of course be translated, the
underlying autograders themselves exist independently of natural
language~\cite{west2021integrating}. Code writing and debugging questions can be
autograded using unit tests. Tracing questions are graded simply by checking if
students entered the correct output. This leaves code comprehension activities,
which require responses to be entered in natural language, as perhaps the only
question in the context of introductory computer science which has historically required
autograders to be built on both a per questions \textit{and} a per language
basis.

\section{Methods}

Our analysis of both the effectiveness of the approach and students' perceptions
and interactions with these problems is divided into two parts. First, in
Section~\ref{subsec:expert_eval}, we present an evaluation of CGBG across a
variety of Indic languages using expert translations of ``gold standard''
English responses to EiPE questions. The second evaluates students' interactions
with and perceptions of EiPL questions in a online National
Programme on Technology Enhanced Learning (NPTEL) course.

\subsection{Evaluation of Expert Translations}\label{subsec:expert_eval}
As a first step towards evaluating the degree to which the Code Generation Based
Grading (CGBG) process enables Explain in Plain Language (EiPL) questions in the
context of Indic languages, we evaluate the model's performance on ten of the most
commonly spoken languages from India's 22 scheduled languages. As shown in
Figure~\ref{fig:languages-by-state}, by using this subset of languages, we cover
a significant proportion of individuals for whom one of these languages is their
mother tongue, according to the 2011 census~\cite{chandramouli2011census}. As
such, this represents the most conservative estimation of the proportion of
speakers for each state who are proficient in one of these languages as we do
not consider individuals who are proficient in multiple languages.

To perform this evaluation we begin with a set of CS1 level code segments in
both C ($n=5$) and Python ($n=6$). Associated with each of these is an example of an
English prompt that successfully generates functionally equivalent code and thus would be graded as `correct' according to CGBG (Table~\ref{tab:questions_examples}). We
invited participants with expertise in a range of Indic languages, recruited through convenience sampling, to translate
these English descriptions of the code into the language(s) in which they self
identified as being fluent. To obtain good-quality translations of these English
descriptions in each of our target Indian languages, we distributed a survey to
Computer Science faculty participants in Faculty Development Programmes (FDPs)
for CS1 courses. Four such FDPs were led by one of the authors, and were
conducted at institutions in Tamil Nadu, Karnataka, Maharashtra, and
Uttarakhand. Prior to distributing the survey, we introduced faculty to EiPE
questions as a way to assess code comprehension, and we described our study of
EiPL questions for linguistically diverse student populations. Faculty
translated English descriptions of the code into the language(s) in which they
self identified as being fluent.\footnote{Faculty were encouraged to preserve technical terms in English if they felt this was more natural, and were asked to transliterate non-English words in Latin script.} 
We then applied CGBG, using GPT-4o as our LLM of
choice, to responses in each of the languages \textit{post hoc} to determine how
well the grading approach performs with respect to each language. Code was
generated using the following prompt:
\begin{tcolorbox}[colback=white, colframe=black, boxrule=0.5pt]
\quoting[leftmargin=0.5pt, rightmargin=0.5pt]
{
Generate a Python function called `foo' that accomplishes the given task using the following instructions written in [SPECIFIED LANGUAGE]: [EXPERT RESPONSE]. The code should be returned in the following format:
\begin{verbatim}
def foo(<parameters here>):
    <code>
\end{verbatim}

Note: The function should always return a value rather than print the result. Additionally, generate only the code and no additional test cases or explanatory text.
}
\end{tcolorbox}

Here [SPECIFIED LANGUAGE] was replaced with the language the expert was
translating the prompt into and [EXPERT RESPONSE] was replaced with the
translation of the prompt. 

\subsection{Evaluating EiPL in an Online Course}\label{}

\begin{figure}[tb]
  \centering
  \includegraphics[width=0.95\columnwidth]{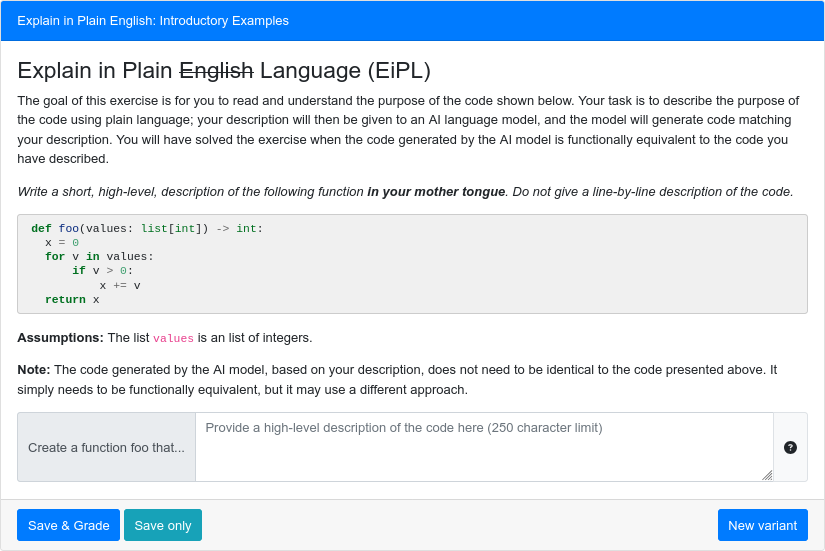}
  \caption{An example of an EiPL question hosted on PrairieLearn.}\label{fig:eipl-question}
\end{figure}

Eight EiPL questions (\autoref{fig:eipl-question}) were then deployed in a Python
programming course hosted by the NPTEL platform\footnote{\url{https://elearn.nptel.ac.in/shop/iit-workshops/completed/programming-with-ai-for-problem-solving/}}. This was an
accelerated course that aimed to teach students the fundamentals of programming
with Generative AI in four weeks of instruction. The instructions of the EIPL
questions alternated between directing students to respond in English and in
their mother tongue. These questions were administered on the PrairieLearn
platform~\cite{west2021integrating} and students were provided with 20 attempts to answer each
question.

\section{RQ1: Success of Expert Translations}

\begin{table*}[t]
    \centering
    \caption{Success of expert translations for each language. The table highlights the total scores and success rates across all questions, with cells color-coded based on performance: \textcolor{good!25}{\fcolorbox{black}{good!25}{\rule{1mm}{1mm}}} purple (75-100\%), \textcolor{okay!25}{\fcolorbox{black}{okay!25}{\rule{1mm}{1mm}}} light blue (50-75\%), \textcolor{notgood!25}{\fcolorbox{black}{notgood!25}{\rule{1mm}{1mm}}} green (25-50\%), \textcolor{bad!25}{\fcolorbox{black}{bad!25}{\rule{1mm}{1mm}}} yellow (0-25\%), and \textcolor{gray!20}{\fcolorbox{black}{gray!20}{\rule{1mm}{1mm}}} grey indicate no responses.}
    \label{tab:questions_examples}
    \resizebox{\textwidth}{!}{
    \large
    \begin{tabular}{>{\raggedright\arraybackslash}p{.5cm} 
    >{\raggedright\arraybackslash}p{4.25cm} 
    >{\raggedright\arraybackslash}p{1.5cm} 
    >{\raggedright\arraybackslash}p{2.2cm} 
    >{\raggedright\arraybackslash}p{2cm} 
    >{\raggedright\arraybackslash}p{2cm} 
    >{\raggedright\arraybackslash}p{2cm} 
    >{\raggedright\arraybackslash}p{2.2cm} 
    >{\raggedright\arraybackslash}p{2cm} 
    >{\raggedright\arraybackslash}p{2.2cm} 
    >{\raggedright\arraybackslash}p{2cm} 
    >{\raggedright\arraybackslash}p{2.2cm}}
        \toprule
        \textbf{} & \textbf{QID} & \textbf{Gujarati} & \textbf{Hindi} & \textbf{Punjabi} & \textbf{Marathi} & \textbf{Bengali} & \textbf{Telugu} & \textbf{Urdu} & \textbf{Kannada} & \textbf{Odia} & \textbf{Tamil} \\
        \toprule
        \multirow{9}{*}{C} & Reverse a String  & \cellcolor{good!25}2/2 (100\%) & \cellcolor{good!25}56/58 (96.6\%) & \cellcolor{good!25}4/4 (100\%) & \cellcolor{good!25}6/6 (100\%) & \cellcolor{good!25}8/9 (88.9\%) & \cellcolor{good!25}68/74 (91.9\%) & \cellcolor{good!25}4/4 (100\%) & \cellcolor{good!25}13/15 (86.7\%) & \cellcolor{good!25}4/4 (100\%) & \cellcolor{okay!25}36/54 (66.7\%) \\ \cline{2-12}
          & Check for Vowel Presence & \cellcolor{good!25}1/1 (100\%) & \cellcolor{okay!25}39/53 (73.6\%) & \cellcolor{good!25}3/3 (100\%) & \cellcolor{good!25}4/5 (80.0\%) & \cellcolor{good!25}6/7 (85.7\%) & \cellcolor{okay!25}41/71 (57.8\%) & \cellcolor{good!25}5/5 (100\%) & \cellcolor{okay!25}9/13 (69.2\%) & \cellcolor{okay!25}3/4 (75.0\%) & \cellcolor{good!25}43/53 (81.1\%) \\ \cline{2-12}
          & Count Even Nums   & \cellcolor{gray!20}0/0 (N/A) & \cellcolor{okay!25}39/52 (75.0\%) & \cellcolor{good!25}1/1 (100\%) & \cellcolor{okay!25}3/5 (60.0\%) & \cellcolor{notgood!25}5/8 (62.5\%) & \cellcolor{okay!25}43/71 (60.6\%) & \cellcolor{good!25}4/4 (100\%) & \cellcolor{okay!25}10/14 (71.4\%) & \cellcolor{good!25}2/2 (100\%) & \cellcolor{notgood!25}27/52 (51.9\%) \\ \cline{2-12}
          & Find Last Zero Index & \cellcolor{gray!20}0/0 (N/A) & \cellcolor{good!25}40/45 (88.9\%) & \cellcolor{good!25}1/1 (100\%) & \cellcolor{okay!25}4/5 (80.0\%) & \cellcolor{good!25}7/7 (100\%) & \cellcolor{okay!25}41/68 (60.3\%) & \cellcolor{okay!25}3/4 (75.0\%) & \cellcolor{okay!25}13/16 (81.3\%) & \cellcolor{good!25}2/2 (100\%) & \cellcolor{okay!25}31/51 (60.8\%) \\ \cline{2-12}
          & Sum of Positive Numbers   & \cellcolor{gray!20}0/0 (N/A) & \cellcolor{good!25}44/47 (93.6\%) & \cellcolor{good!25}1/1 (100\%) & \cellcolor{okay!25}3/4 (75.0\%) & \cellcolor{good!25}7/7 (100\%) & \cellcolor{notgood!25}36/66 (54.6\%) & \cellcolor{notgood!25}2/4 (50.0\%) & \cellcolor{okay!25}10/13 (76.9\%) & \cellcolor{good!25}1/1 (100\%) & \cellcolor{notgood!25}26/52 (50.0\%) \\ 
        \midrule
        \multirow{10}{*}{Py} & Prime Num Check       & \cellcolor{gray!20}0/0 (N/A) & \cellcolor{good!25}41/42 (97.6\%) & \cellcolor{good!25}1/1 (100\%) & \cellcolor{okay!25}3/4 (75.0\%) & \cellcolor{good!25}7/7 (100\%) & \cellcolor{good!25}60/65 (92.31\%) & \cellcolor{good!25}4/4 (100\%) & \cellcolor{good!25}14/14 (100\%) & \cellcolor{good!25}1/1 (100\%) & \cellcolor{notgood!25}28/50 (56\%) \\ \cline{2-12}
                  & Generate Fibonacci List   & \cellcolor{gray!20}0/0 (N/A) & \cellcolor{good!25}35/40 (87.5\%) & \cellcolor{good!25}1/1 (100\%) & \cellcolor{okay!25}2/3 (66.7\%) & \cellcolor{good!25}6/6 (100\%) & \cellcolor{good!25}59/64 (92.2\%) & \cellcolor{okay!25}2/3 (66.7\%) & \cellcolor{good!25}9/11 (81.8\%) & \cellcolor{good!25}2/2 (100\%) & \cellcolor{good!25}41/48 (85.4\%) \\ \cline{2-12}
                  & Get All Even Nums        & \cellcolor{gray!20}0/0 (N/A) & \cellcolor{notgood!25}24/37 (64.9\%) & \cellcolor{good!25}1/1 (100\%) & \cellcolor{okay!25}2/3 (66.7\%) & \cellcolor{good!25}6/7 (85.7\%) & \cellcolor{okay!25}46/62 (74.2\%) & \cellcolor{good!25}3/3 (100\%) & \cellcolor{bad!25}5/13 (38.5\%) & \cellcolor{good!25}1/1 (100\%) & \cellcolor{okay!25}27/48 (56.3\%) \\ \cline{2-12}
                  & Get Largest Positive Num & \cellcolor{gray!20}0/0 (N/A) & \cellcolor{good!25}34/39 (87.2\%) & \cellcolor{good!25}1/1 (100\%) & \cellcolor{good!25}3/3 (100\%) & \cellcolor{good!25}7/7 (100\%) & \cellcolor{good!25}54/61 (88.5\%) & \cellcolor{good!25}3/3 (100\%) & \cellcolor{good!25}12/13 (92.3\%) & \cellcolor{good!25}1/1 (100\%) & \cellcolor{good!25}43/47 (91.5\%) \\ \cline{2-12}
                  & Sum Even Nums in 2D Array & \cellcolor{gray!20}0/0 (N/A) & \cellcolor{okay!25}26/36 (72.22\%) & \cellcolor{good!25}1/1 (100\%) & \cellcolor{good!25}3/3 (100\%) & \cellcolor{okay!25}4/6 (66.7\%) & \cellcolor{notgood!25}25/60 (41.7\%) & \cellcolor{okay!25}2/3 (66.7\%) & \cellcolor{notgood!25}5/12 (41.7\%) & \cellcolor{bad!25}0/1\ \ (0.0\%) & \cellcolor{bad!25}13/45 (28.9\%) \\ \cline{2-12}
                  & Check if Substring Exists& \cellcolor{gray!20}0/0 (N/A) & \cellcolor{okay!25}25/33 (75.8\%) & \cellcolor{good!25}1/1 (100\%) & \cellcolor{bad!25}1/3 (33.3\%) & \cellcolor{bad!25}3/7 (42.9\%) & \cellcolor{okay!25}32/58 (55.2\%) & \cellcolor{good!25}3/3 (100\%) & \cellcolor{okay!25}8/12 (66.7\%) & \cellcolor{bad!25}0/1\ \ (0.0\%) & \cellcolor{bad!25}14/45 (31.1\%) \\
        \midrule
        & \textbf{Totals: } & \cellcolor{good!25}3/3 (100\%) & \cellcolor{good!25}403/482 (83.6\%) & \cellcolor{good!25}16/16 (100\%) & \cellcolor{okay!25}34/44 (77.3\%) & \cellcolor{good!25}66/78 (84.6\%) & \cellcolor{okay!25}505/720 (70.1\%) & \cellcolor{good!25}35/40 (87.5\%) & \cellcolor{okay!25}108/146 (74.0\%) & \cellcolor{good!25}17/20 (85.0\%) & \cellcolor{notgood!25}329/545 (60.4\%) \\
        \bottomrule
    \end{tabular}}
\end{table*}

In total, we collected translations from 180 participants. Though the majority
of the responses were in Hindi, Telugu, and Tamil, all languages except for
Gujarati received responses for all 11 questions
(\autoref{tab:questions_examples}). Overall, the model performed well with the
majority of languages achieving above an 80\% correctness rate overall. The one
notable exception to this was Tamil which not only achieved the lowest
correctness rate overall but also had a wide variance in correctness rates
across the questions. This is particularly surprising given the large quantity
of responses acquired for Tamil, but it aligns with prior work by Jordan et
al.~\cite{jordan2024need} which found GPT-3.5 to be poor at generating questions
in Tamil. 

Though less dramatic, there was also a reasonable amount of variance in the
correctness rates for Hindi and Telugu for which we also had a large number of
responses. These results call for further investigation to understand the
nature of these ``false negatives'', particularly when considering the
sensitivity of LLMs to the specific wording of the input prompts
\cite{denny2023conversing}.  Future work should aim to distinguish between those
that are due to the model's limitations and those that are due to ambiguities
introduced through the translation process. The latter of these is particularly
important as it may be used to inform instruction for students who are
responding to EiPL questions in each of the respective languages.

\section{RQ2: Students' Interactions with and Perceptions of EiPL Questions}

\begin{figure}[tb]
  \centering
  \begin{subfigure}{\columnwidth}
    \centering
    \resizebox{0.95\columnwidth}{!}{
      \begin{tikzpicture}
  \begin{axis}[
    ybar,
    width=10cm,
    height=2.5cm,
    ylabel={Count},
    symbolic x coords={
        Assamese,
        Urdu,
        Marathi,
        Punjabi,
        Gujarati,
        Kannada,
        Bengali,
        Telugu,
        English,
        Malayalam,
        Hindi,
        Tamil
    },
    xtick=data,
    nodes near coords,
    nodes near coords align={vertical},
    x axis line style = { opacity = 0 },
    axis y line       = none,
    tickwidth         = 0pt,
    enlarge x limits  = 0.1,
    bar width=0.25cm,
    ylabel style={font=\large, align=center, text=black},
    xlabel style={font=\large, align=center, text=black},
    xticklabel style={font=\small, text=black, rotate=45, anchor=east},
    yticklabel style={font=\small, text=black}
  ]
  \addplot coordinates {
    (Assamese,1)
    (Urdu,1)
    (Marathi,1)
    (Punjabi,1)
    (Gujarati,1)
    (Kannada,4)
    (Bengali,4)
    (Telugu,6)
    (English,7)
    (Malayalam,10)
    (Hindi,15)
    (Tamil,17)
  };
  \end{axis}
\end{tikzpicture}
    }
    \caption{Languages students reported as their mother tongue(s).}
    \label{fig:languages}
  \end{subfigure}
  \hfill
  \begin{subfigure}{\columnwidth}
  \centering
  \resizebox{0.95\columnwidth}{!}{
      \begin{tikzpicture}
  \pgfplotstableread{
    Semester  VeryLow    Low        Neutral     High       VeryHigh
    Q1        0          0          3.12        25.00      71.88
    Q2        0          15.62      10.94       26.56      46.88
  }\frequency

  \begin{axis}[
    scale only axis,
    name=ax1,
    legend cell align=center,
    legend style={at={(2.75,1.25)},anchor=north,draw=none},
    xbar stacked,
    xmin=-100,
    xmax=0,
    try min ticks=3,
    xticklabel style = {font=\scriptsize},
    xticklabel=\pgfmathparse{abs(\tick)}\pgfmathprintnumber{\pgfmathresult}\,$\%$,
    ytick=data,
    yticklabel style={align=center},
    yticklabels={
      English,
      Mother Tongue, 
    },
    enlarge y limits={abs=0.45cm},
    width=65px,
    height=40px
    ]
    \addlegendimage{area legend, fill=bad, postaction={pattern=crosshatch, pattern color=black}}
    \addlegendimage{area legend, fill=notgood, postaction={pattern=grid, pattern color=black}}
    \addlegendimage{area legend, fill=gray, postaction={pattern=north east lines, pattern color=black}}
    \addlegendimage{area legend, fill=okay, postaction={pattern=dots, pattern color=black}}
    \addlegendimage{area legend, fill=good, postaction={pattern=horizontal lines, pattern color=black}}
    \addplot [fill=gray, postaction={pattern=north east lines, pattern color=black}] table [x expr=-(\thisrow{Neutral}/2), meta=Semester ,y expr=\coordindex] {\frequency};
    \addplot [fill=notgood, postaction={pattern=grid, pattern color=black}] table [x expr=-\thisrow{Low}, meta=Semester ,y expr=\coordindex] {\frequency};
    \addplot [fill=bad, postaction={pattern=crosshatch, pattern color=black}] table [x expr=-\thisrow{VeryLow}, meta=Semester, y expr=\coordindex] {\frequency};
    \addlegendentry{Very Low}
    \addlegendentry{Low}
    \addlegendentry{Neutral} 
    \addlegendentry{High}
    \addlegendentry{Very High}
  \end{axis}

  \begin{axis}[
    scale only axis,
    at=(ax1.south east),
    xbar stacked,
    xmin=0,
    xmax=100,
    xticklabel style = {font=\scriptsize},
    try min ticks=3,
    xticklabel=\pgfmathparse{abs(\tick)}\pgfmathprintnumber{\pgfmathresult}\,$\%$,
    ytick=data,
    yticklabels={},
    enlarge y limits={abs=0.45cm},
    width=65px,
    height=40px
    ]
    \addplot [fill=gray, postaction={pattern=north east lines, pattern color=black}] table [x expr=(\thisrow{Neutral}/2), meta=Semester ,y expr=\coordindex] {\frequency};
    \addplot [fill=okay, postaction={pattern=dots, pattern color=black}] table [x=High, meta=Semester ,y expr=\coordindex] {\frequency};
    \addplot [fill=good, postaction={pattern=horizontal lines, pattern color=black}] table [x=VeryHigh, meta=Semester ,y expr=\coordindex] {\frequency};
  \end{axis}
\end{tikzpicture}
    }
    \caption{``How do you rate your proficiency in \textit{reading and
    writing}''}
    \label{fig:lang_fluency}
  \end{subfigure}
  \caption{Languages spoken by students who participated in the activities
  (\autoref{fig:languages}) and their proficiency in reading and writing in
both their mother-tongue as well as English (\autoref{fig:lang_fluency}).}
\end{figure}
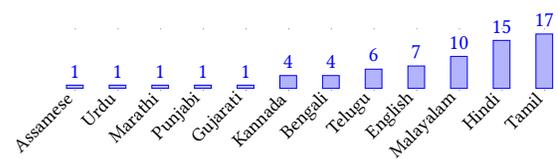
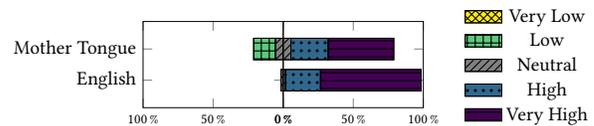

A total of 72 students participated in the study, with 64 providing responses to
the survey. Students were asked to self-report their mother tongue(s) in a
post-activity survey. As shown in \autoref{fig:languages}, participants spoke a
wide variety of languages, notably, with several reporting English as one of their mother
tongues. Students were also asked to report their proficiency in English and
their mother tongue, with the majority reporting high proficiency in both,
although proficiency in their mother tongue was slightly lower
(\autoref{fig:lang_fluency}). Regarding the completion rate of the assignments,
90\% of the students attempted some of the questions, and 65\% completed all of
them.

We report the results of: 1) how students chose to respond to the
EiPL questions and the success of those responses in
Section~\ref{sec:student_interactions} and 2) their perceptions of responding in
their mother tongue in Section~\ref{sec:student_perceptions}.

\subsection{Student Prompts for EiPL Activites}\label{sec:student_interactions}

Though students were requested to respond in their mother tongue for four of the eight 
EiPL activities (due to the request they alternate between languages when answering) we observed that the vast majority of students' final responses
on those questions were in English (78.8\%). While some English responses were
expected given that a number of students reported English as being among their
mother tongues, the proportion of responses in English far exceeds the
proportion of participants which fall under this category. This suggests many of
the participants either had an aversion to, or difficulty with,  responding in
their mother tongue. As for the non-English responses, these often contained a
mix of English and the student's mother tongue. For example,
\begin{itemize}
  \item \textbf{Kannada + English:} ``\textit{s1 mathu s2 anagrams antha check maduvantha function}'' 
  \item \textbf{Hindi + English:} ``\textit{Ek list me kitni bar 0 ata hai wo count krne wala function}''
\end{itemize}

The success rate for students who opted to respond purely in English was 95.9\%
while responses in mixed or pure mother tongue had a success rate of 80\%.
Although we must be cautious about making inferences regarding this
relationship, we make two observations. First, students in both categories
enjoyed reasonable  success suggesting the approach can be effective in
practice. Second, our findings highlight the need for future work to better
understand this disparity, which may be due to deficiencies in the LLM or other
confounding factors e.g., the relationship between English proficiency and
technical competence.

\subsection{Perceptions of Prompting in English vs Mother Tongue}\label{sec:student_perceptions}

\begin{figure}[tb]
  \centering
  \begin{subfigure}{\columnwidth}
    \centering
    \resizebox{0.95\columnwidth}{!}{
      \begin{tikzpicture}
  \pgfplotstableread{
    Semester  VeryLow    Low        Neutral     High       VeryHigh
    Q1        0          0          7.81        37.50      54.69 
    Q2        9.38       18.75      29.69        28.12      14.06
  }\frequency

  \begin{axis}[
    scale only axis,
    name=ax1,
    legend cell align=center,
    legend style={at={(2.75,1.25)},anchor=north,draw=none},
    xbar stacked,
    xmin=-100,
    xmax=0,
    try min ticks=3,
    xticklabel style = {font=\scriptsize},
    xticklabel=\pgfmathparse{abs(\tick)}\pgfmathprintnumber{\pgfmathresult}\,$\%$,
    ytick=data,
    yticklabel style={align=center},
    yticklabels={
      English,
      Mother Tongue, 
    },
    enlarge y limits={abs=0.45cm},
    width=65px,
    height=40px
    ]
    \addlegendimage{area legend, fill=bad!80, postaction={pattern=crosshatch, pattern color=black}}
    \addlegendimage{area legend, fill=notgood!80, postaction={pattern=grid, pattern color=black}}
    \addlegendimage{area legend, fill=gray, postaction={pattern=north east lines, pattern color=black}}
    \addlegendimage{area legend, fill=okay!80, postaction={pattern=dots, pattern color=black}}
    \addlegendimage{area legend, fill=good!80, postaction={pattern=horizontal lines, pattern color=black}}
    \addplot [fill=gray, postaction={pattern=north east lines, pattern color=black}] table [x expr=-(\thisrow{Neutral}/2), meta=Semester ,y expr=\coordindex] {\frequency};
    \addplot [fill=notgood!80, postaction={pattern=grid, pattern color=black}] table [x expr=-\thisrow{Low}, meta=Semester ,y expr=\coordindex] {\frequency};
    \addplot [fill=bad!80, postaction={pattern=crosshatch, pattern color=black}] table [x expr=-\thisrow{VeryLow}, meta=Semester, y expr=\coordindex] {\frequency};
    \addlegendentry{Very Low}
    \addlegendentry{Low}
    \addlegendentry{Neutral} 
    \addlegendentry{High}
    \addlegendentry{Very High}
  \end{axis}

  \begin{axis}[
    scale only axis,
    at=(ax1.south east),
    xbar stacked,
    xmin=0,
    xmax=100,
    xticklabel style = {font=\scriptsize},
    try min ticks=3,
    xticklabel=\pgfmathparse{abs(\tick)}\pgfmathprintnumber{\pgfmathresult}\,$\%$,
    ytick=data,
    yticklabels={},
    enlarge y limits={abs=0.45cm},
    width=65px,
    height=40px
    ]
    \addplot [fill=gray, postaction={pattern=north east lines, pattern color=black}] table [x expr=(\thisrow{Neutral}/2), meta=Semester ,y expr=\coordindex] {\frequency};
    \addplot [fill=okay!80, postaction={pattern=dots, pattern color=black}] table [x=High, meta=Semester ,y expr=\coordindex] {\frequency};
    \addplot [fill=good!80, postaction={pattern=horizontal lines, pattern color=black}] table [x=VeryHigh, meta=Semester ,y expr=\coordindex] {\frequency};
  \end{axis}
\end{tikzpicture}
    }
    \caption{To what extent do you disagree or agree with the statement ``When responding in [given language], this task is an accurate way to evaluate my code comprehension skills?''}
    \label{fig:lang_comp}
  \end{subfigure}
  \hfill
  \begin{subfigure}{\columnwidth}
    \centering
    \resizebox{0.95\columnwidth}{!}{
      \begin{tikzpicture}
  \pgfplotstableread{
    Semester  Low        Neutral     High 
    Q1        0          9.38        90.62
  }\frequency

  \begin{axis}[
    scale only axis,
    name=ax1,
    legend cell align=center,
    legend style={at={(2.75,1.25)},anchor=north,draw=none},
    xbar stacked,
    xmin=-100,
    xmax=0,
    try min ticks=3,
    xticklabel style = {font=\scriptsize},
    xticklabel=\pgfmathparse{abs(\tick)}\pgfmathprintnumber{\pgfmathresult}\,$\%$,
    ytick=data,
    yticklabel style={align=center},
    yticklabels={
        \textcolor{white}{-------------------------}
    },
    enlarge y limits={abs=0.45cm},
    width=65px,
    height=40px
    ]
    \addlegendimage{area legend, fill=notgood, postaction={pattern=grid, pattern color=black}}
    \addlegendimage{area legend, fill=gray, postaction={pattern=north east lines, pattern color=black}}
    \addlegendimage{area legend, fill=okay, postaction={pattern=dots, pattern color=black}}
    \addplot [fill=gray, postaction={pattern=north east lines, pattern color=black}] table [x expr=-(\thisrow{Neutral}/2), meta=Semester ,y expr=\coordindex] {\frequency};
    \addplot [fill=notgood, postaction={pattern=grid, pattern color=black}] table [x expr=-\thisrow{Low}, meta=Semester ,y expr=\coordindex] {\frequency};
    \addlegendentry{Mother Tongue}
    \addlegendentry{Neutral} 
    \addlegendentry{English}
  \end{axis}

  \begin{axis}[
    scale only axis,
    at=(ax1.south east),
    xbar stacked,
    xmin=0,
    xmax=100,
    xticklabel style = {font=\scriptsize},
    try min ticks=3,
    xticklabel=\pgfmathparse{abs(\tick)}\pgfmathprintnumber{\pgfmathresult}\,$\%$,
    ytick=data,
    yticklabels={},
    enlarge y limits={abs=0.45cm},
    width=65px,
    height=40px
  ]
    \addplot [fill=gray, postaction={pattern=north east lines, pattern color=black}] table [x expr=(\thisrow{Neutral}/2), meta=Semester ,y expr=\coordindex] {\frequency};
    \addplot [fill=okay, postaction={pattern=dots, pattern color=black}] table [x=High, meta=Semester ,y expr=\coordindex] {\frequency};
  \end{axis}
\end{tikzpicture}
    }
    \caption{Language preference when responding to EiPL questions}
    \label{fig:lang_pref}
  \end{subfigure}
  \caption{Students' perceptions of the accuracy of the EiPL questions when
  responding in their mother tongue versus English (\autoref{fig:lang_comp})
  and their preference for responding in English or their mother tongue
  (\autoref{fig:lang_pref}).}
\end{figure}
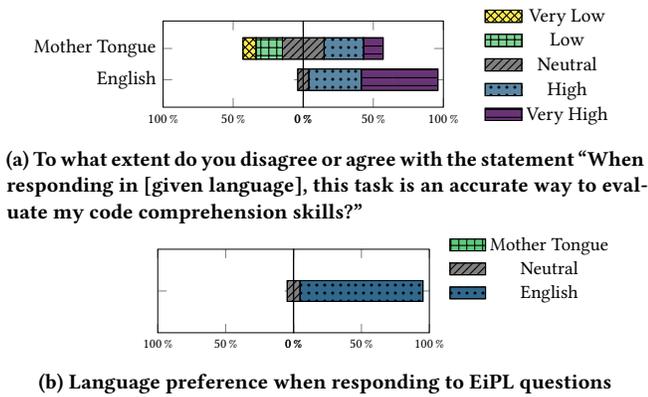

The results of the Likert questions from the survey indicated students were more
inclined to perceive responding in English to be a more accurate assessment of
their code comprehension skills~(\autoref{fig:lang_comp}). Similarly, and in
keeping with the results of the previous section, the vast majority of students
reported that they preferred to respond in English when given the option to
respond in either English or their mother tongue~(\autoref{fig:lang_pref}).

One researcher performed reflexive thematic analysis on the complete set of
responses to the open-ended questions using the guidelines detailed by
Braun and Clarke \cite{braun2006using}. This resulted in three primary themes:
1) students' greater familiarity with English as a technical language, 2)
difficulties writing in their mother tongue, and 3) perceptions of the model as
less capable in their mother tongue.  We discuss each of these themes in detail
below and provide examples of student responses to each.

\paragraph{Students' greater familiarity with English as a technical language}

Many students ($n=10$) reported that they preferred to respond in English as it
was the language they used most frequently when learning or discussing technical subjects.
\begin{itemize}[leftmargin=1em, rightmargin=1em, label={}, itemsep=0.05em, parsep=0.4em]
  \item \emph{``I rarely use my mother tongue with computers. So it has been difficult to explain specific technical ideas effectively in mother tongue. But the same is very easy with English.''} 
\end{itemize}

Several expanded further stating that they used their mother tongue and English
in different contexts, with English being the language they used for technical
subjects and their mother tongue being the language they used in more social 
settings.
\begin{itemize}[leftmargin=1em, rightmargin=1em, label={}, itemsep=0.05em, parsep=0.4em]
  \item \emph{``I am much more fluent in reading and writing in English compared to my mother tongue (Kannada). As I go to an English medium school, I speak, read and write English on a daily basis. However, I speak Kannada only with my family members and have never learnt how to read or write Kannada.''} 
\end{itemize}

\paragraph{Difficulties writing in mother tongue}

Related to the previous section, where students reported that they were more
familiar with English as a technical language and their mother tongue as a
language for general communication, a small number of students ($n=3$) reported 
difficulties writing in their mother tongue.
\begin{itemize}[leftmargin=1em, rightmargin=1em, label={}, itemsep=0.05em, parsep=0.4em]
  \item \emph{``Not able to write in mother tongue but able to only comprehend whereas I have a strong control over English.''} 
  \item \emph{``I am only able to speak in my mother tongue and not able to read or write. I am very proficient in English. Hence, I wouldn't be able to do the task in my mother tongue.''}  
\end{itemize}
These results reflect the results of the Likert question on language proficiency 
discussed in the previous section (\autoref{fig:lang_fluency}).

\paragraph{Perceptions of the model as less capable in mother tongue}

A number of students ($n=7$) reported that they perceived the model to be less 
capable of producing accurate responses when prompted in their mother tongue.
\begin{itemize}[leftmargin=1em, rightmargin=1em, label={}, itemsep=0.05em, parsep=0.4em]
  \item \emph{``I was able to explain my requests well in English and the GPT code functioned well and gave expected results but I wrote the same in plain Hindi, and was struggling to get answers.''} 
  \item \emph{``Sometimes the explanation given in mother tongue is not understood properly by the AI.''} 
\end{itemize}
One reported a specific instance where the model struggled to understand
the prompt when it was in their mother tongue.
\begin{itemize}[leftmargin=1em, rightmargin=1em, label={}, itemsep=0.1em, parsep=0.4em]
  \item \emph{``In English I did not see any issue in comprehending the task by AI, but when I used my mother tongue from one particular question I found that AI unable to convert the `sahi' and `galat' to true and false.''}
\end{itemize}
This may highlight the need for mixing languages (e.g., Tamil + English = ``Tanglish'',
Hindi + English = ``Hinglish'') when responding to EiPL questions  where students use English
keywords when referring to technical concepts and their mother tongue for other
words. It may also be the case that, given many students reported using English
when communicating in a technical context, they were less familiar with
describing technical concepts in their mother tongue.  This may have decreased
the quality of their responses and led to the subsequent lower performance of the approach.

\section{Discussion}

Though the results relating to perceptions of the EiPL questions highlighted
several challenges of deploying these questions in courses, the results of the
expert evaluation are still promising. Additionally, it is important to note
that these results will only improve as both the quality of the LLMs available
and the size of the datasets continue to improve and increase, respectively. As
such, the question of integrating EiPL questions into instruction is less a matter
of ``when'' and more a matter of ``how''.

The results of the classroom study highlight the sensitivity to the local
context when answering the question of ``how''  EiPL questions using CGBG are
deployed. In particular, the qualitative analysis of the survey responses
highlighted the students' greater familiarity with English for describing
programs given its status as a technical \textit{lingua franca}, and issues with
the LLM misinterpreting certain terms from their mother tongue. Though the
former may simply be a byproduct of the population taking the course in which
the study was conducted, the latter is likely to persist across populations.
This may suggest that, when introducing these questions to novices in the
context of Indian education, using a mixed language approach (e.g.,
``Hinglish''~\cite{parshad2016india}) could be beneficial. The primary goal is
to reduce the overhead of the extraneous task of learning English, and to shift
the focus towards learning to comprehend and describe code for populations where
English proficiency is less common. 

Previous work conducted in the context of undergraduate education in Tamil Nadu~\cite{raj2017students,raj2018does} found that students expressed positive
sentiments toward bilingual instruction in Tamil and English. Looking ahead to
the future of EiPL questions, as LLMs improve, they will increasingly enable
multilingual code comprehension questions. As one student noted in the open-ended
survey responses,
\begin{quote}
    ``\textit{Everyone is free to choose their language and it is wonderful if that option is available.}''
\end{quote}
Though our results showed that performance varied between languages, the
fundamental proposition of empowering students to demonstrate code comprehension
in the language of their choosing, thus enabling these questions to be used more
easily in a broader, global context is powerful.

\section{Limitations}

First, there is an inherent difficulty in curating a dataset of expert responses
across multiple languages. Validating the `correctness' of each response
requires significant overhead, as experts in computing education, programming,
and the specific language are needed for each evaluation. The lack of this kind
of validation combined with the use of convenience sampling means that the
quality of responses in the dataset we evaluated may vary across each of the
languages. As such, the results we present should not be interpreted as the
LLM's agreement with ground truth for each of the languages under consideration,
but rather, its agreement with what we can assume are reasonably high quality
translations.

The limitations of the evaluation of the EiPL questions in the NPTEL course are
twofold. First, the accelerated pace of the course (4-weeks) means it likely
attracted individuals already familiar with programming rather than novices.
This may have contributed to an aversion for students to respond in their mother
tongue due to their greater familiarity with using English in a technical
context. Similarly, this was an online course rather than a traditional in
person university course. Future evaluations should consider the use of these
questions in a more traditional university CS1 context.

\section{Conclusions}

Explain in Plain English questions have been a traditional method for developing
and assessing code comprehension skill in introductory computer science courses.
However, challenges with grading these questions have limited their uptake, and
present particular challenges in linguistically diverse regions like India. In
our study, we evaluated the efficacy of the Code Generation Based Grading (CGBG)
approach applied to ten of India's most commonly spoken languages. We
demonstrated that CGBG can effectively generate and grade code from responses in
multiple languages, achieving high correctness rates in most cases. While many
students preferred responding in English due to familiarity and perceived
precision, the approach shows promise for enabling responses in their mother
tongue. Challenges such as difficulties with technical terminology in native
languages and varying model performance were identified, highlighting areas for
future research. This work underscores the potential of Explain in Plain
Language (EiPL) questions to make computer science education more inclusive and
accessible, paving the way for more equitable learning experiences globally.

\balance
\bibliographystyle{ACM-Reference-Format}
\bibliography{refs}

\end{document}